\documentclass{article}
\usepackage{spconf,amsmath,graphicx}
\usepackage{amsfonts, bm}
\usepackage{diagbox,balance}
\usepackage{subcaption,acronym}
\usepackage[dvipsnames]{xcolor}
\usepackage{booktabs}

\acrodef{RTF}{relative transfer function}
\acrodef{LOCA}{local conformal autoencoder}
\acrodef{SLAM}{simultaneous localization and mapping}
\acrodef{TDOA}{time difference of arrival}
\acrodef{DNN}{deep neural network}
\acrodef{RoI}{region of interest}
\acrodef{RIR}{room impulse response}

\acrodef{GCC-PHAT}{generalized cross-correlation}

\title{Unsupervised Acoustic Scene Mapping Based on Acoustic Features and Dimensionality Reduction}
\name{Idan Cohen, Sharon Gannot and Ofir Lindenbaum}
\address{Faculty of Engineering, Bar-Ilan University, Ramat-Gan, 5290002, Israel\\
{\{Idan.Cohen, Sharon.Gannot, Ofir.Lindenbaum}\}@biu.ac.il}
\begin{document}
\ninept
\maketitle

\begin{abstract}
Classical methods for acoustic scene mapping require the estimation of the \ac{TDOA} between microphones. Unfortunately, \ac{TDOA} estimation is very sensitive to reverberation and additive noise.
We introduce an unsupervised data-driven approach that exploits the natural structure of the data.
Toward this goal, we adapt the recently proposed
\textit{\ac{LOCA}} – an offline deep learning scheme for extracting standardized data coordinates from measurements.
Our experimental setup includes a microphone array that measures the transmitted sound source, whose position is unknown, at multiple locations across the acoustic enclosure.
We demonstrate that our proposed scheme learns an isometric representation of the microphones' spatial locations and can perform extrapolation over new and unvisited regions. 
The performance of our method is evaluated using a series of realistic simulations and compared with a classical approach and other dimensionality-reduction schemes. We further assess reverberation's influence on our framework's results and show that it demonstrates considerable robustness.
\end{abstract}

\begin{keywords}
acoustic scene mapping, \ac{RTF}, unsupervised learning,
local conformal autoencoder (LOCA),
dimensionality reduction
\end{keywords}
\section{Introduction}
\label{sec:intro}
From augmented reality to robot autonomy, it is essential to map the environment and the room's shape in many audio applications.
Consequently, there has been growing attention towards the task of \ac{SLAM}. In \ac{SLAM}, a moving observer is considered, following an unknown path, and the task is then to reconstruct both the trajectory and the map of the environment.
Due to the rapid improvement of computer vision technology, visual \ac{SLAM} \cite{durrant2006simultaneous}, based on optical sensors, has been a subject of extensive research. The audio processing community later adopted the same concept, as spatial information can also be extracted from the audio signals \cite{5152813,8340823}.

This work focuses on the environment mapping problem using audio signals acquired by a microphone array, a task closely related to the acoustic \ac{SLAM} problem.
Traditionally, the environment map is specified in terms of landmarks, and reconstruction of the map is equivalent to localizing the landmarks \cite{durrant2006simultaneous}. In our setting, audio signals are emitted from sound sources of unknown positions and recorded by a microphone array that scans the \ac{RoI} and maps it.
Practically, acoustic \ac{SLAM} methods necessitate the use of localization techniques, usually based on \ac{TDOA} estimation between microphone pairs \cite{5152813,8340823}. 
A widely used approach for
\ac{TDOA} estimation is the  \ac{GCC-PHAT} algorithm \cite{Knapp1976}. 
Unfortunately, the performance of \ac{GCC-PHAT} severely degrades in highly reverberated environments \cite{Champagne1996}, resulting in erroneous \ac{TDOA} estimates and poor source localization, especially for large source-sensor distances.
Although many improvements to generalized cross-correlation schemes were proposed, e.g., \cite{599651, dibiase2001robust,Dvorkind2005}, simple \ac{TDOA}-based mappings cannot yield a reliable reconstruction of the environment map.

In this work, we propose to adopt the \acf{RTF} \cite{Gannot2001,markovich2009multichannel} as a feature for the source localization task rather than first estimating the \ac{TDOA}. The \acp{RTF} are high-dimensional acoustic vectors that were shown to lie on manifolds and were successfully applied to audio processing tasks, e.g.~source localization and tracking \cite{LauferGoldshtein2020}, and source separation \cite{markovich2009multichannel}. 
As the \acp{RTF} are lying on a manifold \cite{LauferGoldshtein2020}, they can be naturally integrated with the \ac{LOCA} dimensionality reduction scheme \cite{Peterfreund2020}, enabling the inference of a latent space representation. The learned representation captures the latent variables that parameterize the data, i.e., it can recover the 2-D \ac{RoI}. The manifold assumption has been previously used for audio-based source localization \cite{LauferGoldshtein2020, Deleforge2012, Deleforge2015, LauferGoldshtein2015, lauferGoldshtein2016}, but is now adapted to Acoustic Scene Mapping for the first time.


%
Several key elements reflect the novelty of our approach: 1) We utilize recent progress from the manifold learning field, enabling us to robustly deal with the problem of acoustic scene mapping in reverberant environments circumventing the pre-processing stage of extracting the \ac{TDOA}, which is sensitive to reverberation. Importantly, no prior knowledge about the position of the sound source is required.~2) We introduce a more suitable acoustic feature vector, namely the \ac{RTF}, into the \ac{LOCA} training process.~3) We design a circular microphone array that travels in the \ac{RoI} in a scanner-like manner and is tailor-made for our framework requirements.~4) We change the optimization scheme of \ac{LOCA} to enable stable convergence in our setting.~5) Our method outperforms existing kernel-based schemes in terms of mapping accuracy and time efficiency. Moreover, we show that it can predict the locations of measurements in regions unseen during training, which is usually not the case for more traditional training-based localization schemes. 



\section{Theoretical Background}
\label{sec:background}

\subsection{The \acf{RTF}}
\label{ssec:RTF}
The \acp{RTF} \cite{Gannot2001,markovich2009multichannel} are used as the acoustic feature for our approach. They are independent of the source signal and carry relevant spatial information. Consequently, they can serve as \emph{spatial fingerprints} that characterize the positions of each of the sources in a reverberant enclosure \cite{LauferGoldshtein2020}.

\smallskip
\noindent\textbf{RTF - Definition and Estimation:} 
For a pair of microphones, we consider time-domain acoustic recordings of the form
\begin{equation}
    x_i(n) = \{a_i\ast s\}(n) + u_i(n),
\end{equation}
with $s(n)$ being the source signal, $a_i(n)$, the \acp{RIR} relating the source and each of the microphones, $i = \{1,2\}$,
and $u_i(n)$ noise signals independent of the source.
We define the acoustic transfer functions $A_i(k)$ as
the Fourier transform of the \acp{RIR} $a_i(n)$. Following \cite{Gannot2001}, the \ac{RTF} is defined as $H(k) = \frac{A_2(k)}{A_1(k)}$.
%
Assuming negligible noise and selecting $x_1(n)$ as the reference signal, the \ac{RTF} can be estimated from:
\begin{equation}\label{eq:01}
    \hat{H}(k) = \frac{\hat{S}_{x_2x_1}(k)}{\hat{S}_{x_1x_1}(k)} \approx H(k),
\end{equation}
where $\hat{S}_{x_1x_1}(k)$ and $\hat{S}_{x_2x_1}(k)$ are the power spectral density (PSD) and the cross-PSD, respectively.

\smallskip\noindent\textbf{Representing \acp{RTF} on the Acoustic Manifold:}
Previous works have studied the nature of the RTFs \cite{lauferGoldshtein2016, LauferGoldshtein2015, LauferGoldshtein2020}.
RTFs are high-dimensional representations in $\mathbb{C}^D$ that correspond to the vast amount of reflections from different surfaces characterizing the enclosure.
Following the manifold hypothesis \cite{Tenenbaum2000}, we assume that the RTF vectors, drawn from a specific \ac{RoI} in the enclosure, are not spread uniformly in the entire space of $\mathbb{C}^D$. Instead, they are confined to a compact nonlinear manifold $\mathcal {M} $ of dimension $d$, much smaller than the ambient space dimension, i.e., $d \ll D$.
This assumption is justified by the fact that perturbations of the \ac{RTF} are only influenced by a small set of parameters related to the physical characteristics of the environment. These parameters include the enclosure dimensions, its shape, the surfaces’ materials, and the positions of the microphones and the source. Moreover, we focus on a static configuration, where the enclosure properties and the source position remain fixed.
In such an acoustic environment, the only varying degree of freedom is the location of the microphone array.

Although nonlinear, in small neighborhoods, the manifold \textit{is} locally linear, meaning that it is flat in the vicinity of each point and coincides with the tangent plane to the manifold at that point. Hence, the Euclidean distance can faithfully measure affinities between points that reside close to each other on the manifold. For remote points, however, the Euclidean distance is meaningless.
This property is utilized to our advantage by \ac{LOCA},
which makes use of the linear relation between points in a local neighborhood.
%
\begin{figure}[htbp]								
\vspace{ -0.15 in}
\centering
\includegraphics[width=0.65\columnwidth]{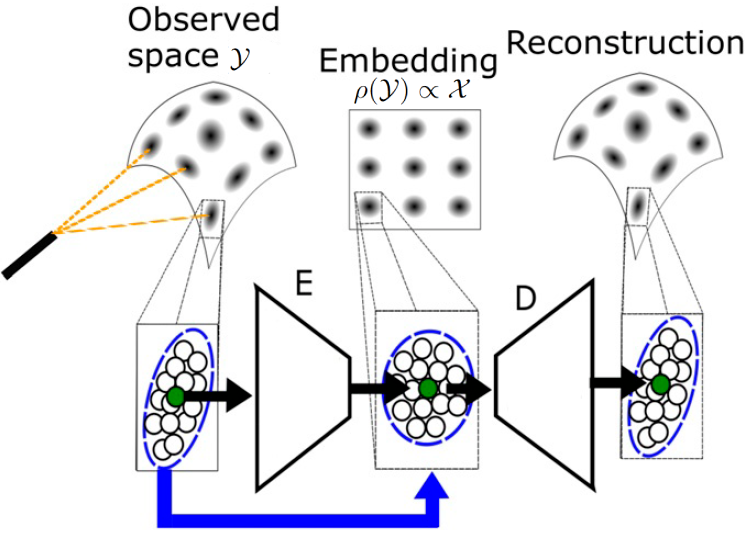}
\caption{An illustration of LOCA adapted to acoustic scene mapping. The observation space $(\mathcal{Y})$ is assumed
to model a nonlinear deformation of the inaccessible manifold $(\mathcal{X})$. We
attempt to invert the unknown measurement function, utilizing
the bursts sampling strategy. LOCA consists of an encoder ($E$) parameterized by $\rho$ and a
decoder ($D$) parameterized by $\gamma$. 
The autoencoder receives a set of points and corresponding neighborhoods; each neighborhood
is depicted as a dark oval point cloud (at the top of the figure), which we implement using a microphone array. At
the bottom, we zoom in onto a single anchor point $y_i$ (green) and its
corresponding neighborhood $Y_i$ (bounded by a blue ellipsoid). The encoder
attempts to whiten each neighborhood in the embedding space, while the
decoder aims at reconstructing the input.
In practice, a dedicated microphone array constellation will be used to extract bursts of adjacent acoustic samples of \acp{RTF} (see Fig.~\ref{fig:setting}).}
\vspace{-0.1 in}
\label{fig:loca_illustration}
\end{figure}
\subsection{Local Conformal Autoencoder (LOCA)}
\label{ssec:LOCA}
\noindent\textbf{Motivation:}
Our approach requires a localized sampling strategy denoted \emph{burst sampling} \cite{Singer2008}.
A burst is a collection comprising samples taken from a local neighborhood.
Consequently, bursts can provide information on the \emph{local} variability in the vicinity of each data point, allowing for estimating
the Jacobian (up to an orthogonal transformation) of the unknown measurement function.

\smallskip\noindent\textbf{Assumptions and Derivation:}
First, for simplicity, we consider the case where the latent domain of our system of interest is a path-connected domain in the Euclidean space $\mathcal{X}\subset\mathbb{R}^d$.
Observations of the system consist of samples captured by a measurement device and are given as a nonlinear smooth and bijective function $f:\mathcal{X}\rightarrow\mathcal{Y}$, where $\mathcal{Y}\subset\mathbb{R}^D$  is the ambient or measurement space, and $D\gg d$.
The high dimensional acoustic features, namely the \acp{RTF}, constitute the measurement space in our setting.   

Consider $N$ \ac{RTF} data points, denoted $\bm{x}_1,\ldots,\bm{x}_N \in
\mathbb{R}^d$ in the latent space. Assume that all these points lie on a path-connected, $d$-dimensional subdomain of $\mathcal{X}$. Importantly, we do not have direct access to the latent space $\mathcal{X}$. Samples in the latent space $\mathcal{X}$, which can be thought of as latent states, are pushed forward to the ambient space $\mathcal{Y}$ via the unknown deformation $f$. 
Let $\{\bm{y}_i\}_{i=1}^N \in \mathbb{R}^D$ be a set of measurements, with $\bm{y}_i=f(\bm{x}_i)$. 
We assume that an observed burst around each $\bm{y}_i$ consists of perturbed versions of the latent state $\bm{x}_i$, pushed through the unknown deformation $f$. Formally, for fixed $1\leq i\leq N$, we define the burst (sample cloud) $\bm{Y}_i=\{\bm{y}_i^{(j)}\}_{j=1}^M$ as the set of independent and identically distributed samples of the random variable
$f(\bm{X}_i) \in \mathbb{R}^D$, where $\bm{X}_i\sim\mathcal{N}(\bm{x}_i, {\sigma}^2\bm{I}_d)$, are independent random variables and $M$ denotes the number of samples in each burst.
The available data now consists of $N$ bursts of observed states and their perturbed samples.
We assume that $\sigma$ is sufficiently small such that, practically, the differential of $f$  does not change within a ball of radius $\sigma$ around any point.
Such sufficiently small $\sigma$ allows us to capture the local neighborhoods of the states at this measurement scale on the latent manifold.

Ideally, being able to find $f^{-1}:\mathcal{Y}\rightarrow\mathcal{X}$ would suffice to reconstruct the latent domain, but even if $f$ is invertible, it is generally not feasible to identify $f^{-1}$ without access to $\mathcal{X}$.
We can, however, try to construct an embedding $\rho:\mathbb{R}^D\rightarrow\mathbb{R}^d $
that maps the observations $\{\bm{y}_i\}\in{\mathcal{Y}}$ so that the image of $\rho\circ f$ is isometric to $\mathcal{X}$ when $\sigma$ is known.
In our Euclidean setting, it is equivalent to require that the distances in the latent space are preserved, i.e., samples should satisfy $\left\lVert \rho(\bm{y}_i) - \rho(\bm{y}_j)\right\rVert_2 = \left\lVert \bm{x}_i -  \bm{x}_j\right\rVert_2$ for any $i,j$.
%
Following the mathematical derivation in \cite{Peterfreund2020}, the embedding $\rho$ that preserves the distances should satisfy the condition
\begin{equation}\label{eq:02}
    \frac{1}{{\sigma}^2}{\bm{C}(\rho(\bm{Y}_i))}=\bm{I},
\end{equation}
where $\bm{C}$ is the covariance matrix computed over the embedding of the samples from each burst.
In other words, \eqref{eq:02} requires whitening the covariance of local samples in the embedding domain.

\smallskip\noindent\textbf{Implementation:}
As stated above, our task is to find the embedding function $\rho$. We use an encoder neural network $(E)$ to learn the embedding function.
Following \eqref{eq:02}, we define the loss term to be minimized at each iteration:
\begin{equation}
    L_{\textrm{white}}(\rho)=\frac{1}{N}\sum_{i=1}^N\left\lVert\frac{1}{{\sigma}^2}{\bm{\hat{C}}(\rho(\bm{Y}_i))}-\bm{I}_d\right\rVert_F^2 ,
\end{equation}
with $\rho$ the learned embedding function up to the current iteration and $\bm{\hat{C}}(\rho(\bm{Y}_i))$ is the empirical covariance over a set of $M$ realizations
$\{\rho(\bm{y}_i^{(j)})\}_{j=1}^M$.
Next, we note that $f$ is an invertible function, and so is its inverse $f^{-1}$. Since our embedding $\rho$ tries to estimate $f^{-1}$ up to an orthogonal transformation, we can enforce the invertibility of $\rho$, reducing ambiguity in the solution.
The invertibility of $\rho$ means that there exists an inverse mapping $\gamma: \mathbb{R}^d\rightarrow\mathcal{Y}$ such that $\bm{y}_i=\gamma(\rho(\bm{y}_i))$ for any $1\leq i\leq N$. By imposing an invertibility property on
$\rho$, we effectively regularize the solution of $\rho$ away from noninvertible
functions.
Practically, we impose invertibility by using a decoder neural network $(D)$ parameterized by $\gamma$.
This approach induces a loss term based on the mean square error (MSE) between samples and their reconstructed counterparts:
\begin{equation}
    L_{\textrm{recon}}(\rho,\gamma)=\frac{1}{N\cdot{M}}\sum_{i,m=1}^{N,M}\left\lVert \bm{y}_i^{(m)}-\gamma(\rho(\bm{y}_i^{(m)}))\right\rVert_2^2.
\end{equation}
The two-loss terms are equally weighted. The \ac{LOCA} framework is schematically depicted in Fig.~\ref{fig:loca_illustration}.

\begin{figure}[!t]								
\vspace{ -0.15 in}
\centering
\includegraphics[width=0.6\columnwidth]{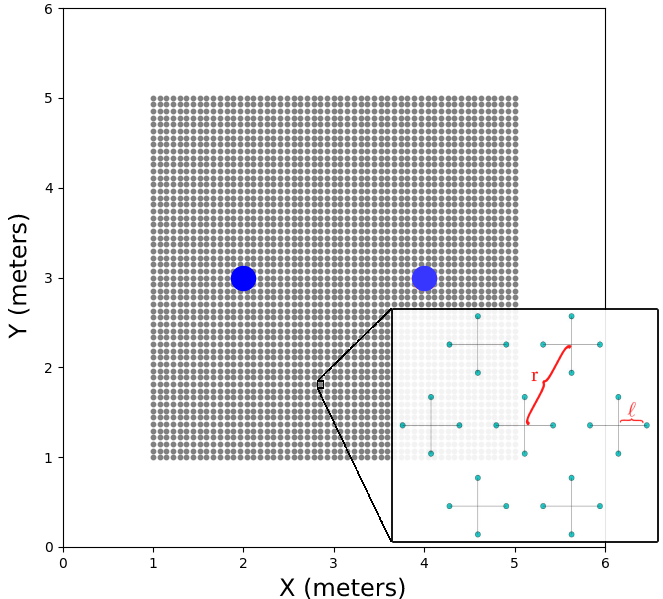}
\caption{Room sampling strategy visualization. Blue circles denote the location of the sound sources. The sampling grid along which the device travels is shown in grey. We zoom in to show the schematic configuration of the burst microphone array: seven cross-like components consisting of vertical and horizontal microphone pairs.}
\label{fig:setting}
\vspace{ -0.1 in}
\end{figure}

\section{Simulation Setup}\label{sec:simulation}
\noindent\textbf{Experimental Setup:}
To implement the burst sampling strategy, we construct a microphone array with seven cross-shaped sub-arrays in a circular constellation, each consisting of four omnidirectional microphones. The distance between the microphones in each arm of the cross is $2\ell$, and the radius w.r.t.~the constellation's center is $r$. 
For a relatively small radius, this circular array accurately simulates a Gaussian neighborhood sampled in the latent domain, as the burst sampling strategy requires.
Practically, $\ell$ might be larger than $r$, resulting in overlapping crosses of microphones (see Fig.~\ref{fig:setting}).

Consider a room of dimensions $[6,6,2.4]$~m. Define a \ac{RoI}  within the room of size $4\times4$~m, which is symmetrically positioned around the center of the room, $0.2$~m above the room's floor.
The following simulation will be divided into two parts: a learning phase and a test phase.
In the learning phase, the microphone array is free to move only on a grid with a spatial resolution of $56$ samples along each dimension, yielding a total of $N=3136$ bursts for training and validation, with a spacing of $7.27$~cm between the centers of neighboring bursts.
In the test phase, we allow the device to travel in the same confined region but not necessarily on the grid. The test set includes 750 bursts from random locations in the \ac{RoI}.
Two omni-directional sound sources are placed at  $[x,y,z]=[2,3,1.7]$~m and $[x,y,z]=[4,3,1.7]$~m, lying $1.5$~m above the sampling-grid plane. Note that the positions of the sound sources are unknown to the algorithm.
The height of the sound sources above the sampling plane is vital due to simple geometric considerations. It allows phase accumulation between received signals relative to the sources in all directions.
In addition, it became apparent by a series of experiments that using only a single emitting source may produce heavily deformed embeddings at the edges far from the source. For that reason, we used two sources in the simulations. Note that the sound sources are not simultaneously active, thus allowing for the estimation of the individual \acp{RTF}.
A top-view visualization of the sampling constellation is shown in Fig.~\ref{fig:setting}.


\smallskip\noindent\textbf{Reverberated Data and Training:}
The reverberant acoustic data was generated using the GPU-accelerated \ac{RIR} Simulator \cite{DiazGuerra2020, allen1979image}. For the learning-phase data, the synthetic \acp{RIR}, $a_i(n)$, were convolved with $5$ second long white noise signals, $s(n)$. The acoustic data at the test phase is simulated by convolving the \acp{RIR}, $a_i(n)$, with $8$ second long speech signals, $s(n)$, drawn from the TIMIT dataset \cite{AB2_SWVENO_1993}. We examined three reverberation times ($\textrm{RT}_{60}=160, 360, 610 $~ms), set the sampling frequency to $f_s=16$~kHz, and the speed of sound to $c=343$m/s.
We also set the microphone array parameters to $r=2$~cm and $\ell=3$~cm.
Using this configuration, at any given location and for each source, we can estimate the horizontal \acp{RTF} and the vertical \acp{RTF}, having a total of seven horizontal and seven vertical \acp{RTF} for a single burst.
\ac{RTF} estimation was applied according to \eqref{eq:01}, using $N_{\textrm{FFT}}=256$, with 50\% overlapping Hamming windows.
A single estimated \ac{RTF} is a complex-valued feature of length $\frac{N_{\textrm{FFT}}}{2}+1=129$. For working with real-valued \acp{DNN}, we concatenated the real and imaginary parts to form a 258-dimension real-valued feature. 
%
It turns out that picking a portion of the RTF bins is preferable. We obtain the best results when taking bins 5 to 99, corresponding to frequencies $312.5$--$6190$~Hz.

Our data tensor is constructed as follows: we take 95 bins from each complex \ac{RTF} for each single sound source to construct a real-valued feature of length 190. Since we have vertical and horizontal \ac{RTF} components in our burst, we concatenate them to obtain a feature vector of length 380. Finally, having two speech sources, we repeat the process and concatenate both results. We eventually end up with a data tensor of shape $[N, M, D]=[3136, 7, 760]$.
In terms of the LOCA terminology, we can write that $\mathcal{Y} \subset \mathbb{R}^{760}$, and we wish to reconstruct a 2-D latent embedding, $\mathcal{X} \subset \mathbb{R}^{2}$.




\begin{figure}[htbp]
\vspace{ -0.1 in}
\addtolength{\belowcaptionskip}{5pt}
\addtolength{\abovecaptionskip}{5pt}
\begin{subfigure}[b]{0.32\linewidth}
 \includegraphics[width=1\textwidth]{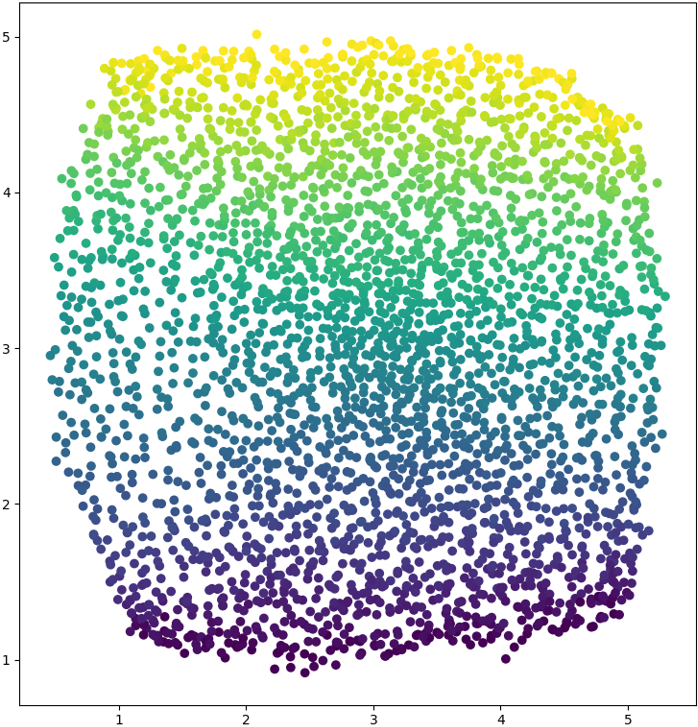}
 \caption{$\textrm{RT}_{60}=160$~ms}
 \label{fig:emb160}
\end{subfigure}
\begin{subfigure}[b]{0.32\linewidth}
 \includegraphics[width=1\textwidth]{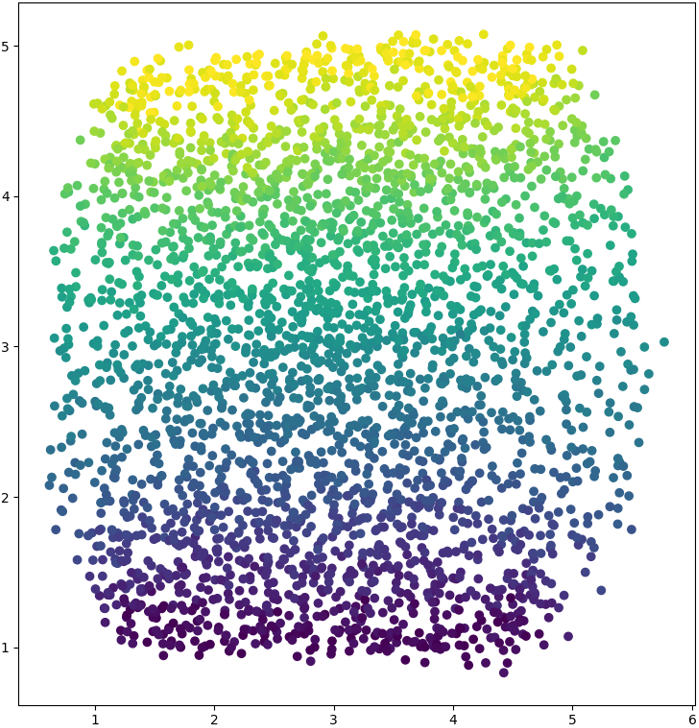}
 \caption{$\textrm{RT}_{60}=360$~ms}
 \label{fig:emb360-1}
\end{subfigure}
\centering
\begin{subfigure}[b]{0.32\linewidth}
 \includegraphics[width=1\textwidth]{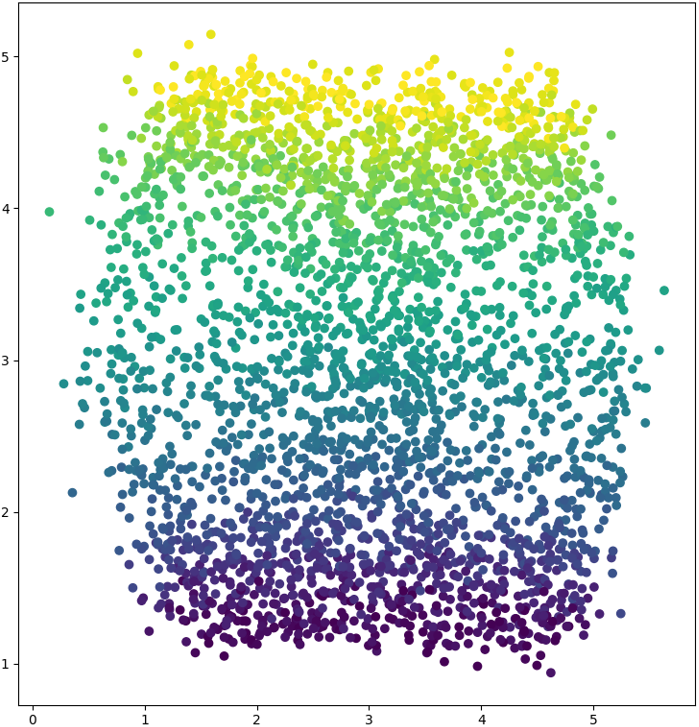}
 \caption{$\textrm{RT}_{60}=610$~ms}
 \label{fig:emb360-2}
\end{subfigure}
\caption{2-D geometric reconstruction achieved by the embedding of our framework (LOCA). The coloring indicates the correlation with the true vertical coordinate of the original scene.
}
\vspace{ -0.07 in}
\label{fig:emb_results}
\end{figure}

\smallskip\noindent\textbf{Training and Parameters:}
We implemented both the encoder and decoder using simple feed-forward architectures. The encoder consists of an initial layer of dimension $D=760$ with no nonlinearity, followed by five layers of size $200$ with `Leaky-Relu' activations, ending with a latent representation layer of size $d=2$. Similarly, the decoder consists of the latent representation layer with no nonlinearity, followed by four layers of size $200$ with `tanh' activations, an additional linear layer of size $200$, and an output reconstruction layer of size $D$.
We use a batch size of $2048$ samples and a learning rate of $10^{-3}$.
While the original \ac{LOCA} formulation applied alternating minimization steps between the whitening and reconstruction loss terms, we found that, in our setting, a more stable solution is obtained by minimizing a combined loss function.
%
We use $90\%$ of the learning-phase data during training, and the rest $10\%$ are used for validation and determining the best model weights.
Finally, based on the geometry of the microphone array, we set $\sigma=6\cdot 10^{-4}$.


\section{Simulation Study and Analysis}
\label{sec:sim_study_analysis}
As discussed earlier, our embeddings are invariant to shifts and orthogonal transformations; therefore, they should be calibrated using a set of anchor points. Typically, three anchor points are required to recover the required transformation. Note that the calibration process is the first time we incorporate our knowledge of the true geometry of the problem into the model. In Fig.~\ref{fig:emb_results}, we visualize the calibrated embedding provided by our framework.
Our learned embedding demonstrates a high correlation between the main directions of the embedding and the true $x-y$ axes characterizing the square sampling grid.
While increasing the reverberation level may deform the embedding, the model can still capture the directional correlation and reveal the square shape of the grid (see Fig.~\ref{fig:emb_results}).
We suspect the reverberation-attributed degradation stems from the \ac{LOCA}'s assumption that $f$ is a smooth function. 
In high reverberation, small perturbations in the location translate to relatively sharp changes in the \ac{RTF}, leading to less smooth $f$.
As the high-frequency bins tend to change rapidly with the burst's position, it was empirically verified that we should discard these bins to achieve better results.

\begin{table}[htbp]
\addtolength{\belowcaptionskip}{6pt}
\addtolength{\abovecaptionskip}{-0pt}
\caption{MAE (cm) of reconstruction using test data.}
\centering
\begin{tabular}{lccc}
 \toprule
 Method & $160$~ms & $360$~ms & $610$~ms\\
 \midrule
 PCA & 72.2 & 72.5 & 74.3 \\
 MMDS & 73.4  & 79.1 & 82.5 \\
 DM & 20.4 & 22.7 & 65.1 \\
 A-DM & 17.8 & 18.3 & 33.8 \\
 {LOCA} & $\mathbf{11.3}$ & $\mathbf{13.4}$ &  $\mathbf{18.5}$  \\
 \bottomrule
\end{tabular}
\vspace{-0.12 in}
\label{table:1}
\end{table}

\smallskip\noindent\textbf{Comparison with Manifold Learning Methods:}
Next, we compare the results with other baseline methods. Since we focus on unsupervised acoustic scene mapping, we compare LOCA to alternative dimensionality reduction methods.
Specifically, we compare to: Principal-Component-Analysis (PCA), Metric-Multi-Dimensional-Scaling (MMDS) \cite{abdi2007metric}, diffusion maps (DM) \cite{coifman2006diffusion}, and anisotropic diffusion maps (A-DM) \cite{Singer2008}.
The above methods are commonly used in manifold learning applications and are justified by our assumption of the existence of the acoustic manifold. Hence, they may serve as valid candidates for comparison.
We stress that these schemes are tested with the same data used to test LOCA.
 
As discussed earlier, we quantify the performance using the test-phase data. We calculate the mean absolute error (MAE) of the distance between the positions in the embedding space and their matching counterparts in the latent space. Note that we first calibrate the embedding using an orthogonal transformation and a shift before calculating the MAE. Additionally, in DM and A-DM, where we need to carry out hyper-parameter tuning for $\sigma$, we first find the optimal $\sigma$ value by selecting the embedding that leads to the minimal MAE, given a small set of four anchors.
In Table~\ref{table:1}, we compare the performance measures of all methods for the different reverberation times.
It is evident that LOCA leads to the best results in terms of MAE for all reverberation levels.
Finally, it is important to note the advantage of LOCA over the other methods regarding inference time. Once a mapping of the \ac{RoI} has been constructed in the training phase, we can perform self-localization using new samples, where now we can use only a single cross instead of an entire burst. LOCA presents a fast and simple inference process based on the \ac{DNN}'s forward pass. In contrast, Kernel-based methods require calculating the kernels between the new and all other training samples, which is a cumbersome process. To quantify this gap, we compared the inference time of our method with kernel schemes and observed that we could reduce the run time by four orders of magnitude.

To complete the study, we have tested a na\"{i}ve classical approach for acoustic scene mapping of the discussed \ac{RoI}. We considered a coarser grid than before, consisting of $10\times10$ samples, with a spacing of 44.4~cm between neighboring locations. To further simplify, the sound sources were lowered to the grid plane. For this approach, which is further explained in Section 3.6 of \cite{Dorfan2020}, we used two crosses of microphone pairs identical to the central cross array in Fig.~\ref{fig:setting}. One is a static reference array located at the middle of the \ac{RoI}, whereas the second array is successively moved along the grid points. We then used \ac{GCC-PHAT} to estimate the \acp{TDOA} between the moving array and the reference array centroid w.r.t.~each speaker. Additionally, we used SRP-PHAT \cite{do2007real} to estimate the DOA of each speaker w.r.t.~each array. This scheme lead to MAEs of 13.7~cm for $\textrm{RT}_{60}=160$~ms and to 74.8~cm for $\textrm{RT}_{60}=610$~ms. For 
this evaluation grid \ac{LOCA}'s performance remains relatively stable.

\begin{figure}[htbp]
\addtolength{\abovecaptionskip}{14pt}
 \addtolength{\belowcaptionskip}{-10pt}
 \vspace{-0.25 in}
    \begin{subfigure}[t]{0.33\linewidth}
        \includegraphics[width=1\textwidth]{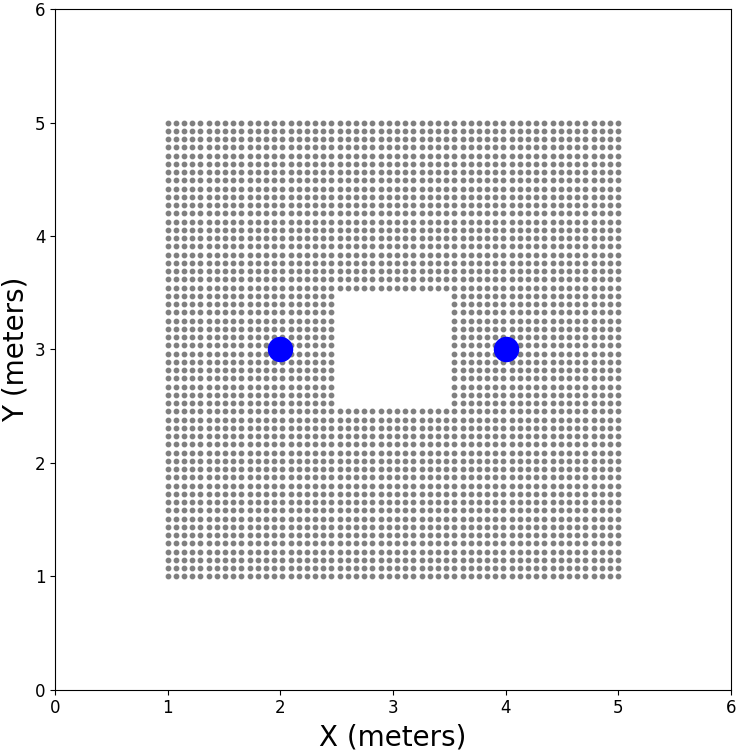}
        \caption{Training region}
        \label{fig:interp:a}
    \end{subfigure}%
    \begin{subfigure}[t]{0.33\linewidth}
        \includegraphics[width=1\textwidth]{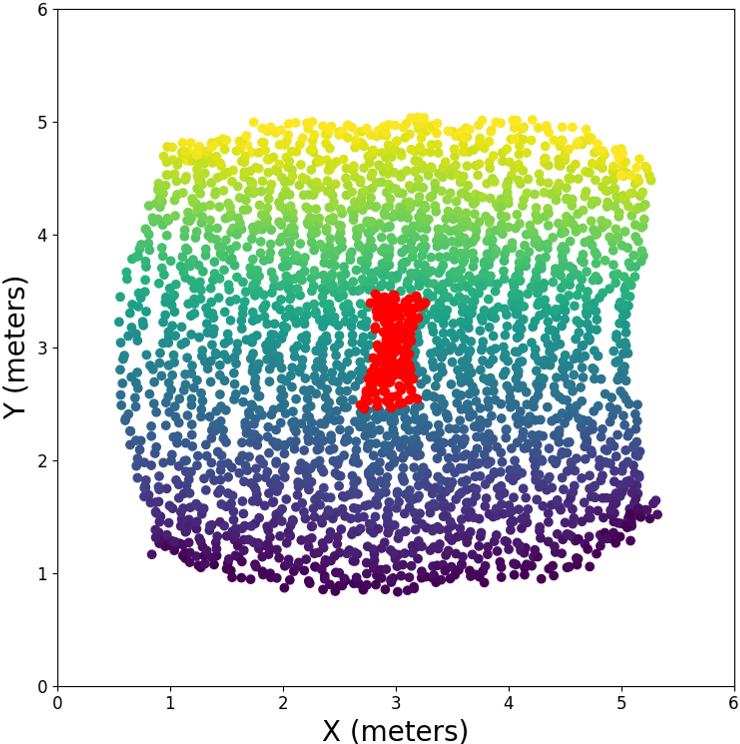}
        \caption{LOCA}
        \label{fig:interp:b}
    \end{subfigure}
    \begin{subfigure}[t]{0.33\linewidth}
        \includegraphics[width=1\textwidth]{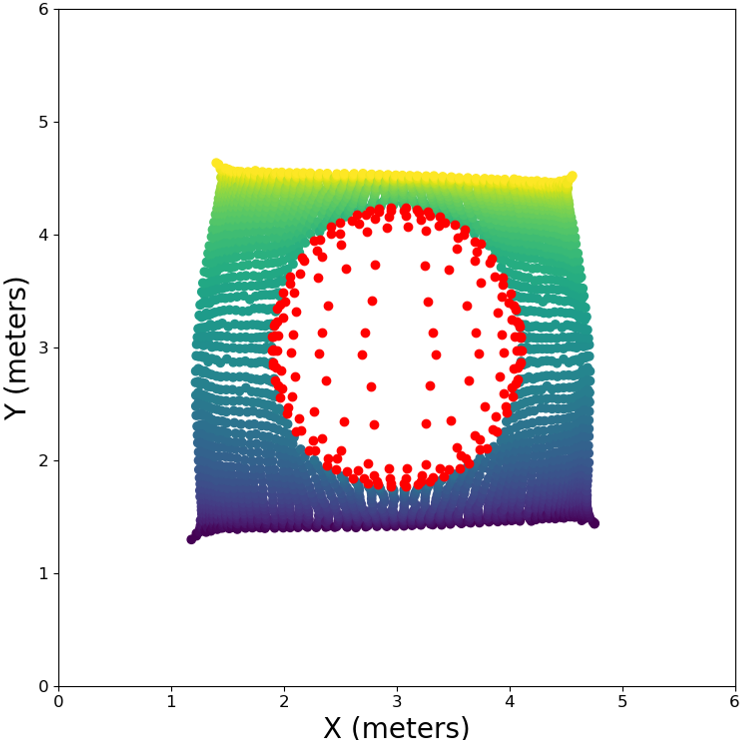}
        \caption{\centering A-DM}
        \label{fig:interp:c}
    \end{subfigure}
    \caption{A visualization of the extrapolation capabilities of LOCA and A-DM for $\textrm{RT}_{60}=160$~ms. The samples from the extrapolated region are shown in red. The MAE values of the extrapolated region are $16.1$~cm for LOCA and $67.4$~cm for A-DM.}
 \vspace{-0.1 in}
    \label{fig:interp}
\end{figure}

\smallskip\noindent\textbf{Extrapolation Capabilities:}
 We now examine the extrapolation capabilities of the proposed scheme and compare them to the A-DM manifold learning scheme. For that, we modify our simulation setup as follows. The manifold is learned using the $[1,5]\times[1,5]$~m \ac{RoI}, excluding the $[2.5,3.5]\times[2.5,3.5]$~m region in the center of the room, as depicted in Fig.~\ref{fig:interp:a}. We only demonstrate the performance for the $\textrm{RT}_{60}=160$~ms case. As evident from Figs.~\ref{fig:interp:b},~\ref{fig:interp:c} \ac{LOCA} extrapolates very well to the excluded region compared to the A-DM.

\section{Conclusions}
In this work, we harness recent advances in the field of \ac{DNN}-based manifold learning to deal with the problem of acoustic scene mapping in a reverberant environment.
We demonstrate the applicability of the \ac{RTF} as a proper feature vector, which 
encapsulates the relevant spatial information for the task at hand.
We apply an approach that utilizes the natural structure of the data and the local relations between samples, avoiding the severe performance degradation typical to other localization schemes.
Our simulation results demonstrate the ability to perform extrapolation to regions not seen during training while significantly reducing run-time requirements compared with existing schemes.
Furthermore, we confirm the robustness of the approach even in mild to high reverberation.



\balance
\bibliographystyle{IEEEbib}
\bibliography{refs}

\end{document}